\documentclass[a4paper,10pt,twoside]{cpc-hepnp}

\usepackage{multicol}
\usepackage{graphicx}
\usepackage{amssymb,bm,mathrsfs,bbm,amscd}
\usepackage[tbtags]{amsmath}
\usepackage{lastpage}

\def\cs137 {$^{137}$Cs~}
\def\co60 {$^{60}$Co~}
\def\eu152 {$^{152}$Eu~}
\def\gam {$\gamma$~}
\def\kcps {$\times 10^3 s^{-1}$~}
\def\cps {$ s^{-1}$~}

\begin{document}


\title{Conditioning the $\gamma$ spectrometer for activity measurement at very high background
\thanks{Supported by the National Science and Technology Major Project under Grant No. ZX06901, by National Natural Science
Foundation of China under Grant No. 10975083 and by Tsinghua
University Initiative Scientific Research Program } }
\author{%
YAN Weihua , ZHANG Li-Guo, ZHANG Zhao, XIAO
Zhi-Gang\email{xiaozg@tsinghua.edu.cn}}

\maketitle

\address{
(Department of Physics, Tsinghua University, Beijing 100084)}
\address{
(Institute of Nuclear Energy and Technology, Tsinghua University,
Beijing 100084)}

\maketitle


\begin{abstract}
The application of a high purity Germanium (HPGe) $\gamma$
spectrometer in determining the fuel element  burnup in a future
reactor is studied. The HPGe detector is exposed  by  a \co60 source
with varying  irradiation rate from 10 \kcps to 150 \kcps to
simulate the input counting rate in real reactor environment. A
\cs137 and a \eu152 source are positioned at given distances  to
generate certain event rate in the detector with the former being
proposed as a labeling nuclide to measure the burnup of fuel
element.  It is shown that both the energy resolution slightly
increasing with the irradiation rate and the passthrough rate at
high irradiation level match the requirement of the real
application. The influence of the background is studied in different
parameter sets used in the
 particularly developed procedure of the background subtraction.  It is
demonstrated that with the typical input irradiation rate and \cs137
intensity relevant to deep burnup situation, the precision of the
\cs137 counting  rate in the current experiment is consistently
below 2.8\%, indicating a promising feasibility of utilizing an HPGe
detector in the burnup measurement in future bed-like reactor.

\end{abstract}
\begin{keyword}
fuel element burnup,  $\gamma $ activity,   HPGe, $\gamma$
spectrometer

\end{keyword}
\begin{pacs}
29.30.Kv 28.41.Bm 28.50.Dr
\end{pacs}

\section {Introduction}
The fuel balls  undergo a multipass cycle in the modular pebble bed
reactors (MPBR). Because of the serious inaccuracy of the
computational method, which is conventional for performing the
in-core fuel management in the existing water reactor,  a
non-destructive determination of the burnup is desired to provide
the distributed controlling system  with an online
circulation/discharge judgement on a pebble-by-pebble basis in some
bed-like reactors\cite{HAW02,BSU06}. Gamma ray spectrometry has been
proposed as an effective non-destructive method to determine the
burnup \cite{MAT95,MAT97,LTE00,ANS07,WIL06} as well as the spatial
irradiation distribution \cite{KHA10,MATS75} by measuring the
activity (activity ratio) of given monitor nuclide(s). Among all
kinds of fission products, \cs137 has been proposed to be one of the
effective burnup indicators for its long life-time and the clear
correspondence to the burnup of the fuel pebble with rather
resistance to the power history \cite{ZLG08,FRE06}. Nevertheless,
the fission of the fissile materials in MBPR produces various
radioactive nuclides that emit amount of \gam peaks over the whole
spectrum, and hence, as shown by various simulation results
\cite{HAW02,HAW05,CHEN03,ZLG09} and some testing results, the \cs137
full energy peak is likely obscured by the neighboring \gam rays in
a typically short cooling time. For instance, in almost all the
stage of the burnup, the peak of 658 keV from $^{97}$Nb is
presented. Thus, in order to identify the $\gamma$ peak from the
\cs137 662 keV, a \gam spectrometer with energy resolution of better
than 2 keV is highly desired. In this regard, the High Purification
Germanium (HPGe) \gam detector has been widely proposed in the
determination of fuel pebble burnup as a non-destructive method.

Another feature of the fuel element  with a certain burnup in the
certain future reactor is the high \gam radioactivity. According to
the prediction, the radiation exposing to the HPGe detector varies
from tens of \kcps to more than 150 \kcps, depending on the
shielding condition, the geometric collimation and the cooling time.
In the design of the spectrometer, although the dead time of the
spectrometer can be lowered by decreasing  the irradiation level
through a strict geometrical collimator, the net count rate of the
labeling \cs137 is also suppressed and a larger statistical error is
expected. Therefore for the  MPBR under design, it is of extreme
importance to optimize the working parameters, including the shaping
time, geometrical collimation and shielding, to have the activity
determined with a minimized experimental error limit within the
typical real time (tens of seconds) after very short cooling time
(typically tens of hours).

In this paper, we present the main results in experimental
conditioning of an HPGe \gam spectrometer. The passthrough  curves
at different shaping time and flattop  are measured. For the
appropriately chosen shaping time, the peak resolution is measured
with various radiation background levels. Furthermore, the precision
of measuring the net counts of a \cs137 source is studied in detail
by varying the real time, the net rate of the nuclide of interest
(NOI) and the radiation background level. In Section II, the
experimental setup and conditions are described. Section III
presents the main results and Section IV is the summary.

\section {The experimental setup}
The whole spectrometer consists of an HPGe $\gamma$ detector, an
electrical cooling device (X-Cooler II) and a suitable fast
electronics (DSPEC+) from ORTEC. The cylindric HPGe crystal is 43 mm
in height and 62.6 mm in diameter, with a $\phi 10.8 mm\times 30 mm$
hollow copper electrical pole fed in the axial center. The
efficiency of the HPGe detector for \co60 $\gamma$ rays is about
30\% at the working high voltage +2300V.  The preamplifier adopts
the light feed-back technique to reduce the pile-up effect at high
incident rate.

Three \gam sources are used in the experiment.  A \co60 is applied
to simulate the  irradiation background from 10 to 150 \kcps (dubbed
from CO1 to CO15 in the text) by varying its distance to the
detector.  The intense incident counting rate is controlled by the
\co60 source throughout the test unless specially specified.  A
\cs137 is placed at two positions in the vicinity of the detector to
generate about 30 (CS1) or 120 (CS2) \cps counting rate in the
detector to mimic the counting rate of a low and a high burnup
element, respectively. Correspondingly a \eu152 source is also
introduced to investigate the performance of the spectrometer
varying with the \gam energy. At each geometrical configuration, the
measurement is done over 2 settings of the shaping time and the
flattop (SHP1 and SHP2)  of the main amplifier for 1 or 10 groups,
with 5 files in each group. Totally 2904 files are acquired. With
these measurements it is possible to study the repeatability and the
statistic fluctuation of multi observables.   Table \ref{conditions}
summarizes the experimental conditions of the multi source
experiment.

\begin{table}[h]
\caption { Conditions in the multi source experiment}
\label{conditions}
\begin{center}
\begin{tabular}{ c c c c c c c c } \hline\hline

\multicolumn{1}{c}{}&\multicolumn{2}{c}{Shaping time $\tau_s$  }              &\multicolumn{2}{c}{\cs137 Intensity $n_{Cs}$ (\cps)}&\multicolumn{3}{c}{\co60 Irradiation } \\
\multicolumn{1}{c}{}& \multicolumn{2}{c}{and flattop  $\tau_f$ ($
\mu s$) }& \multicolumn{2}{c}{  and real time $T_r$  (s) }
&\multicolumn{3}{c}{rate $n_{Co}$ (\kcps)}        \\ \hline
   Subscript             &  SHP1    &    SHP2                                         &  CS1       & CS2                           &     CO1       & ...          & CO15  \\  \hline
   Parameter   &      $\tau_s=0.8$ &   $\tau_s=1.2 $     & $ n_{Cs}=30 $      & $ n_{Cs}=130 $     &     $n_{Co}=10$       &...          & $n_{Co}=150$  \\
   Setting        &       $\tau_f=0.8$  &   $\tau_f=  0.6 $    & $ T_{r}=10 $      & $ T_{r}=25 $              &                                &          &     \\ \hline\hline

\end{tabular}
\end{center}
\end{table}

\section {Results and discussions}

\subsection{Passthrough  and energy resolution }

\begin{figure}[h]
\centering
\includegraphics[angle=0,width=0.45\textwidth]{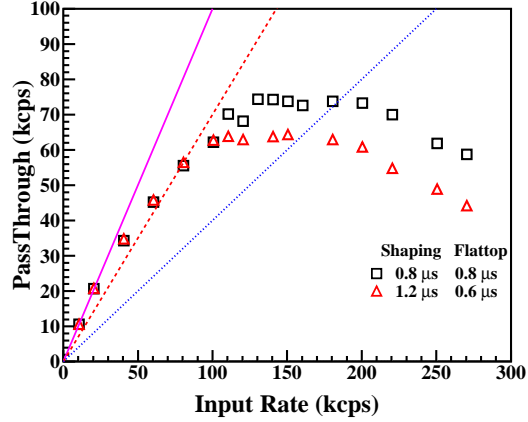}
\caption{(Color online) Passthrough rate as a function of incident
counting rate with SHP1 and SHP2. The solid, dashed and dotted lines
represent certain dead time of 0\%, 30\% and 60\%, respectively.}
\label{pass}
\end{figure}

Figure \ref{pass} presents the passthrough curves measured with SHP1
and SHP2. The abscissa depicts the incident $\gamma$ counting rate
characterizing the radiation background level exposed on the
detector. The low threshold is 0.3 keV. It is shown that at low
radiation level, the passthrough rate increases linearly with the
input rate. At about 100 \kcps, however, the passthrough starts to
deviate with different shaping time and undergoes a plateau up to an
incident rate of about 200 \kcps. Above 200 \kcps, the passthrough
starts to decrease with the input counting rate due to the rapidly
increasing dead time. Unlike the situation with low input rate,
where the passthrough shows insignificant dependence on the shaping
time, the maximum passthrough on the plateau decreases with
increasing the shaping time and flattop as shown in the figure.
Quantitatively the passthrough is 10 \kcps higher with SHP1 than
that with SHP2. In the real application, the amplifier parameters
and the geometrical collimation are so designed that the system is
running on the passthrough plateau to acquire the highest statistics
while keeping adequate energy resolution since the \gam energy peaks
in the proximity of the \cs137 indicator enrich. As shown later, one
can not switch the shaping time down to very small value for the
rapidly degraded energy resolution.

\begin{figure}[h]
\centering
\includegraphics[angle=0,width=0.45\textwidth]{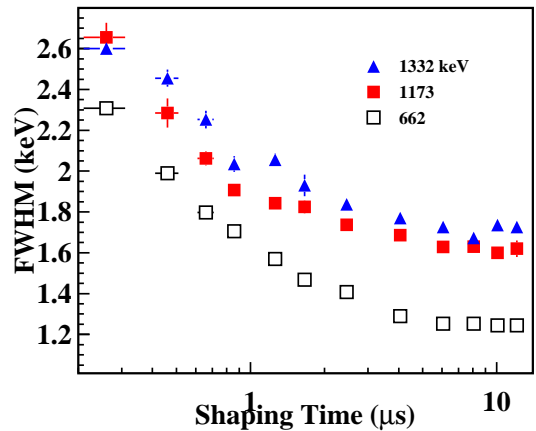}
\caption{(Color online) FWHM of the full energy peaks as a function
of shaping time for the \cs137 peak  (662 keV) and the \co60 peaks
(1173 and 1332 keV). } \label{fwhm1}
\end{figure}

Figure \ref{fwhm1}  depicts the dependence of FWHM on the shaping
time for \cs137  and \co60 peaks. With increasing the shaping time,
the FWHM of the full energy peaks decreases rapidity below about 4
$\mu$s and then saturates at 1.3 keV and 1.8 keV for \cs137 and
\co60 full energy peaks, respectively. Shaping time of larger than
0.8 $\mu$s is preferred with respect to the required resolution of
1.8 keV for $^{137}$Cs. But because of the high expectation of
passthrough, optimization is required. In this paper, only the
results with SHP1 and SHP2 are relevantly presented since these two
conditions meet the requirements of the design of \gam spectrometry
for the future real application.

\begin{figure}[h]
\centering
\includegraphics[angle=0,width=0.45\textwidth]{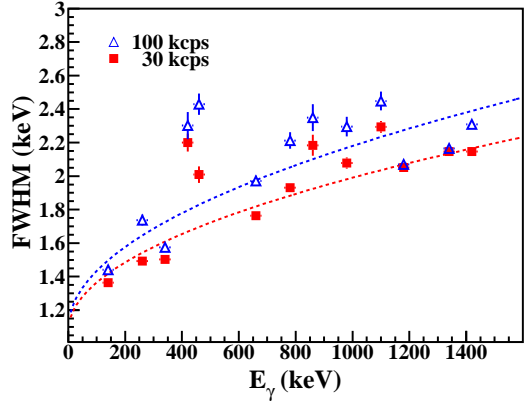}
\caption{(Color online) FWHM of the full energy peaks as a function of peak energy at different input rate.  }
\label{fwhm2}
\end{figure}

The energy dependence of FWHM is plotted in Figure \ref{fwhm2} at
two typical radiation background levels 30 and 100 \kcps. The data
are fitted with a formulae of $a\sqrt{E_{\gamma}+b}$  as shown by
the dashed curves.  Although on average the data points follow the
curves well at both irradiation rates, rather fluctuation of the
data points is visible.  This is attributed to the relative
intensities at different energies of the \eu152 peaks. The full
energy peaks with lower intensities, for instance, at
$E_{\gamma}=411 $ or 1090 keV, suffer more from the background level
induced by the Compton plateau of the intense \co60 source. And
equivalently, the FWHMs become smaller if the  incident rate is
lowered to 30 \kcps as the Compton background decreases. As
expected, the FWHMs of the two \co60 peaks at 1173 and 1332 keV are
not changed by the radiation rate.

\begin{figure}[h]
\centering
\includegraphics[angle=0,width=0.45\textwidth]{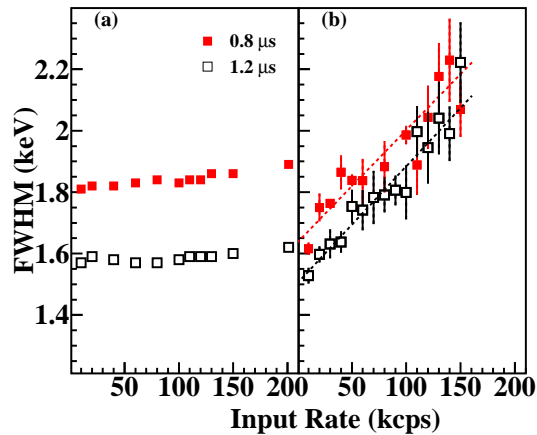}
\caption{(Color online) FWHM of the full energy peak of \cs137 as a
function of  input counting rate at different shaping time for the
situations where the incident counting rate is controlled by \cs137
itself (a) and by the \co60 source (b).} \label{fwhm3}
\end{figure}

Figure \ref{fwhm3} presents the FWHM of the \cs137 full energy peak
as a  function of incident counting rate at different shaping time
configuration SHP1 and SHP2, respectively. Specially the irradiation
rate corresponding to the left panel is controlled by moving the
\cs137 itself. In this case the FWHM keeps constant with the input
counting rate. However, in the test where the input counting rate is
controlled by moving the \co60 while the \cs137 position is fixed,
the FWHM exhibits an increasing trend with the input counting rate.
At typical input rate of 30 \kcps corresponding to deep burnup, the
FWHM of \cs137 full energy peak is below 1.8 keV. The energy
resolution of \cs137 peak is degraded by less than 20\% at 150 \kcps
input rate.

\subsection{Precision of the net counts of the NOI }

The relationship between the  activity of the source $A$ and the net
counting  rate of the full energy peak $n$ recorded by the detector
is written as
\begin{equation}
n=A \Gamma \varepsilon
\end{equation}

where $\varepsilon$ is the full energy detection efficiency and
$\Gamma$ is the absolute strength of the peak.  $n$ is obtained by
subtracting the background $B$ from the total counting $T$  in the
range of interest (ROI) corresponding to the full energy peak range
via
\begin{equation}
n=\frac{N}{t}=\frac{T-B}{t}
\end{equation}

where  $N$ is the net counts recorded in the live time $t$ of
measurement. The standard deviation of the net count $\sigma_N$
writes
\begin{equation}
\sigma_N=  \sqrt{(\sigma_T)^2+(\sigma_B)^2}=\sqrt{N+2B}
\label {sigN}
\end{equation}

In real application, the total background $B$ is derived by
extrapolating the integrated counts $B_w$ in two predefined windows
(with $CH_b$ channels for each) at the board  to the whole ROI (with
$2CH_{hr}$ channels in total) via
\begin{equation}
B=B_w\frac{CH_{hr}}{CH_b}
\label {sigN2}
\end{equation}

\begin{figure}[h]
\centering
\includegraphics[angle=0,width=0.45\textwidth]{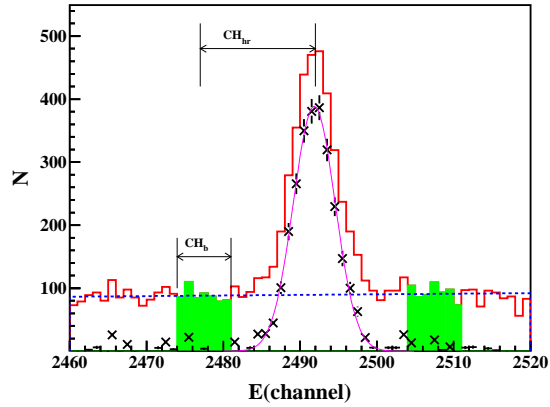}
\caption{(Color online) Procedure of the background subtraction. The
histogram shows the original measured \gam spectrum, and the green
shadowing area depicts the windows  used to fit the background with
$CH_b$ channels for each. The distance between each broad of the ROI
to the peak is denoted as $CH_{hr}$. The cross symbols represent the
net peak after subtracting the fitted background denoted with the
dashed line, and the pink curve is a Gaussian fit to the net peak. }
\label{method1}
\end{figure}

It is found in the analysis that the net counting rate depends
slightly on the range of ROI for the same spectrum. In order to
minimize the fluctuation caused by the change of ROI range, we fix
the number of channels of the ROI for each given energy peak at a
certain shaping time. Figure \ref{method1} shows the sketch of the
procedure of the background subtraction. On both sides of the ROI,
each of which is $CH_{hr}$ to the center of the peak,  a fixed
window with $CH_b$ channels at the broad is selected to derive the
background by a linear fitting to the  data points in these two
windows (the green shadowed area). The net peak (the cross symbols)
is consequently obtained by subtracting the fitting background(the
dashed line) from the total spectrum (the solid histogram). The net
area of the peak is then obtained by fitting the net peak with a
Gaussian function (the solid curve).

\begin{figure}[h]
\centering
\includegraphics[angle=0,width=0.45\textwidth]{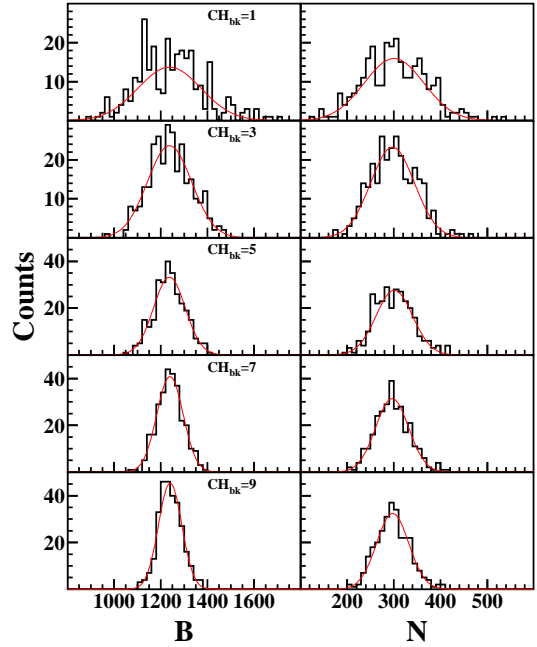}
\caption{(Color online) Distributions of the background $B$ (left)
and the net counts of \cs137 peak $N$  with different number of
channels $CH_b$ of the background window (see Figure \ref{method1})
used to subtract the background} \label{bk1}
\end{figure}

Writing the averaged count of the background in a single channel as
$\hat b$ for a given radiation background level, from Formulae
(\ref{sigN}), one then gets the uncertainty of net area

\begin{equation}
\sigma_N=\sqrt{T+\sigma_{B_{w}}^2\frac{CH_{hr}^2}{CH_{b}^2}}=\sqrt{N +B_w\frac{CH_{hr}}{CH_{b}} +B_w\frac{CH_{hr}^2}{CH_{b}^2}}
=\sqrt{N+\hat b CH_{hr}(1+ CH_{hr}/CH_{b}) }
\label {sigN3}
\end{equation}
It is shown that the uncertainty of the net area or net rate depends
on the background level $\hat b$ as well as  the net area $N$. Under
a given radiation condition, the contribution of the background to
$\sigma_N$ decreases with the width of the background window because
the fluctuation of the background level, shown with the dashed line
in Figure \ref{method1}, is weakened by averaging on more channels,
as indicated by Formulae (\ref{sigN3}). Figure \ref {bk1} presents
the distribution of the background (the left panels) and the net
area (the right panels) of \cs137 peak at different background
window width $CH_b$ for CS1 at about 30 \kcps incident counting
rate. With $CH_b$ tuning from 1 to 9, the distributions of the net
counts and the background become narrower, indicating a decreasing
standard deviation  with $CH_b$.

\begin{figure}[h]
\centering
\includegraphics[angle=0,width=0.45\textwidth]{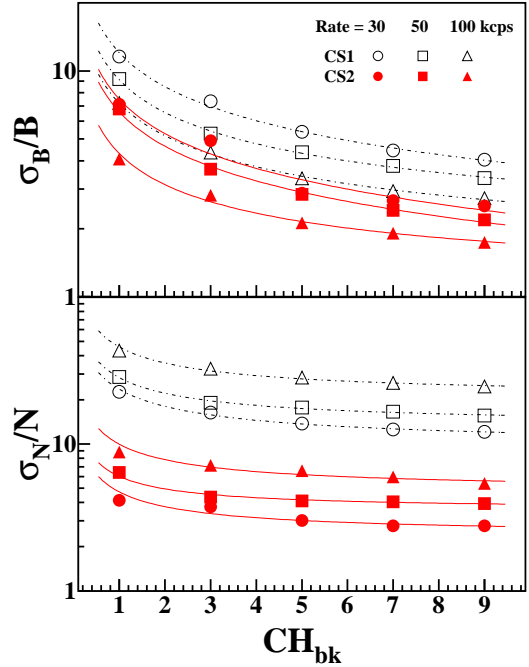}
\caption{(Color online) The relative standard deviation of the
background $B$ (left) and of the net rate of \cs137 peak $n$ as
 a function of  $CH_b$ for CS1 ad CS2 at different input counting rates.}
\label{bk2}
\end{figure}

 Figure \ref {bk2}  presents quantitatively the relative standard deviation of the background
 $\sigma_B/B$ (upper)  and of the  net area $\sigma_N/N$ (lower)
 of \cs137 peak as a function of $CH_b$ for
 CS1 and CS2 at 30, 50 and 100 \kcps incident counting rate, respectively.
 The curves are the fitting with formulae $\sqrt{c_1+c_2/N_b}$ with $c_1$ and $c_2$ being two fitting parameters.
 The relative standard deviation of the background  is higher at lower irradiation rate while the one of the net area decreases
 with lowering the radiation background level. The effect of counting time is also clear by comparing the group of CS1 and CS2.
 With longer measuring time (CS2), both the background and the net area exhibit less uncertainty.  At the typical radiation
 level of 30 \kcps incident rate for the deep burnup fuel element with a \cs137 counting rate at about 130 \cps,
 the precision of the net rate is better than 3\%.

\begin{figure}[h]
\centering
\includegraphics[angle=0,width=0.45\textwidth]{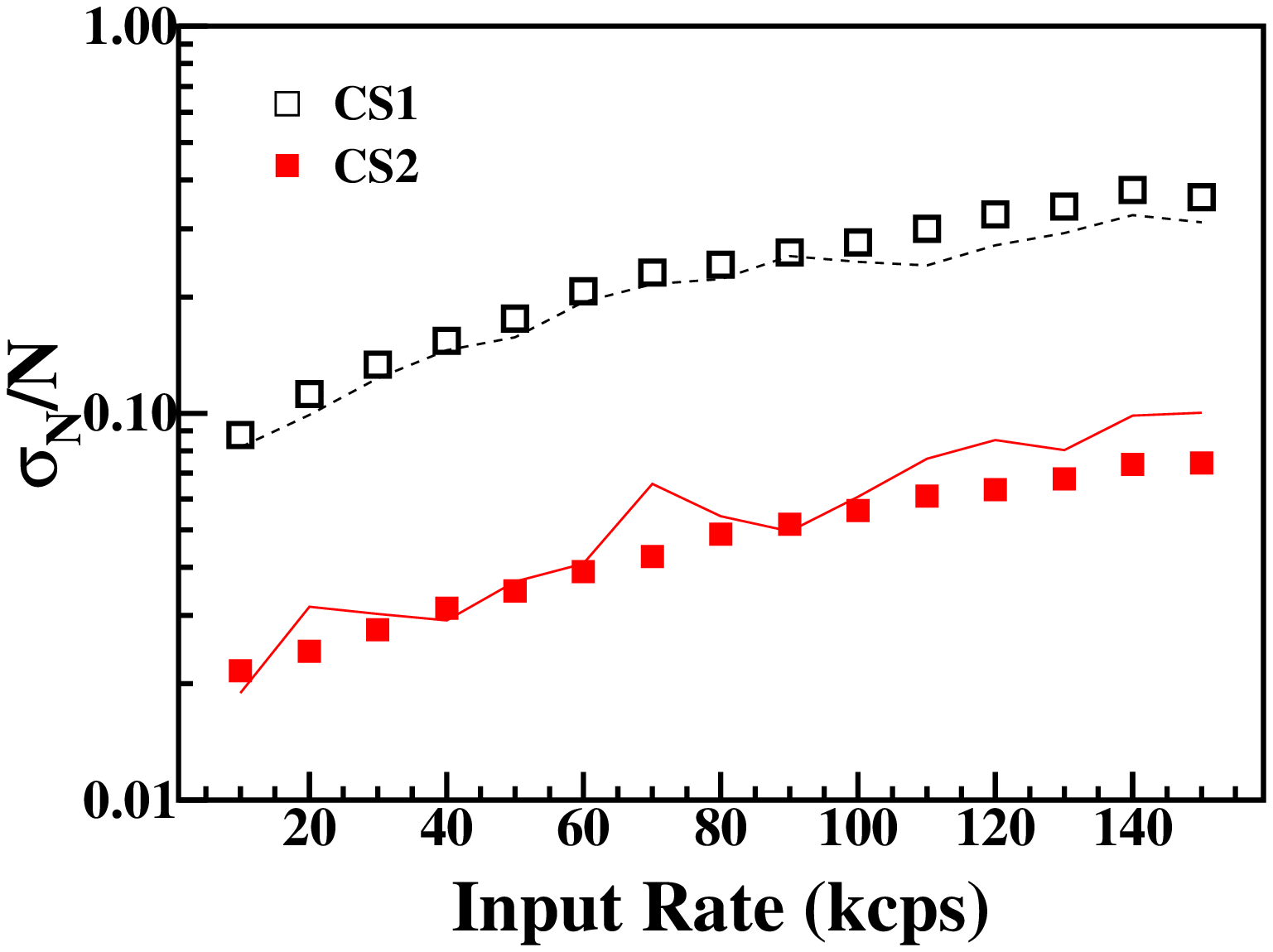}
\caption{(Color online) Relative uncertainty of the net area as a
function of \co60 irradiation rate for CS1 and CS2.  The curves
denote the relative stand deviation of the net area distributions,
and the symbols is  calculated from Formulae (\ref{sigN}) . }
\label{error1}
\end{figure}

Figure \ref{error1} shows the  relative standard deviation profile
of the net area of \cs137 peak as a function of \co60  irradiation
rate for CS1 and CS2, respectively. The symbols  are  calculated via
Formulae (\ref {sigN}) with the total area $T$ and the background $B$.
The curves , on the other hand, are derived from the corresponding
root mean square (RMS) of the net area distributions, as shown
exemplarily in the right columns of Figure \ref{bk1} at the
corresponding radiation level by $RMS/N$. It is shown that the
curves follow the symbols well with few exceptions where the
measurements are repeated for less than 5 times.  It is shown that
the uncertainty of $N$ increases near linearly with the radiation
background level, showing similar trends for CS1 and CS2. For the
group of CS2, for which the \cs137 source is closer to the detector
and the real time is longer, the uncertainty is much smaller.

\begin{figure}[h]
\centering
\includegraphics[angle=0,width=0.6\textwidth]{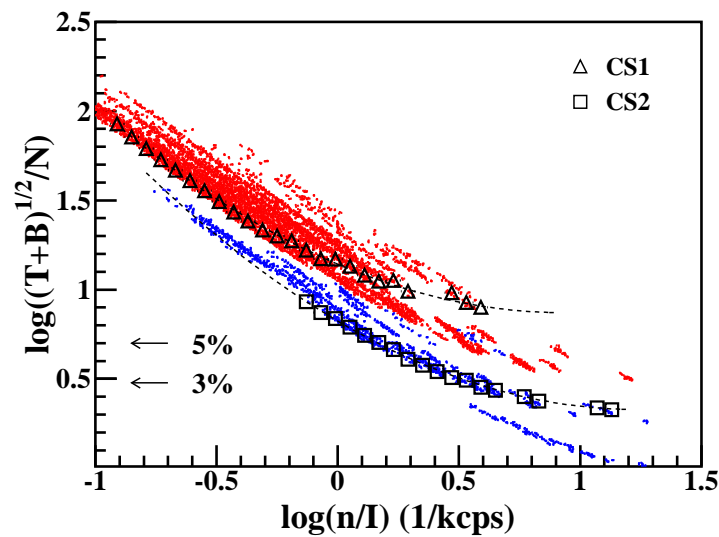}
\caption{(Color online) Correlation between the relative uncertainty
of the net rate and the net rate scaled with the irradiation rates}
\label{error2}
\end{figure}

More investigations including the behavior of  \eu152 peaks in the
same measurements reveal more information on the influence of the
background level.  Figure \ref {error2}  shows the correlation
between the relative standard deviation for all energy peaks below
\co60 1173 keV as a function of the ratio of the net rate and the
irradiation rate. Again the relative uncertainty is calculated via
Formulae (\ref{sigN2}). The pink area corresponds to the group CS1
and the red area to the group CS2.  It is clear that by increasing
the measure time and moving the source closer to the detector, the
overall uncertainties are reduced. Since the background level
induced by \co60 varies with the energy and the net rate differs
among the \eu152 peaks, the uncertainty exhibits a rather broadening
at a given scaled net rate $n/I$. The open triangles and the open
squares with quadratic fitting (the dashed curves) present the
evolution of the uncertainty of \cs137 peak for CS1 and CS2,
respectively. From this curve we can estimate the precision of
measuring the \cs137 net rate if knowing roughly the ratio of the
net counting rate and radiation background level. By neglecting the
detailed distribution  of the background sources, a precision of
2.8\% is again estimated for the deep burnup element in 25-second
measurement in the MPBR application under design.

\begin{figure}[h]
\centering
\includegraphics[angle=0,width=0.6\textwidth]{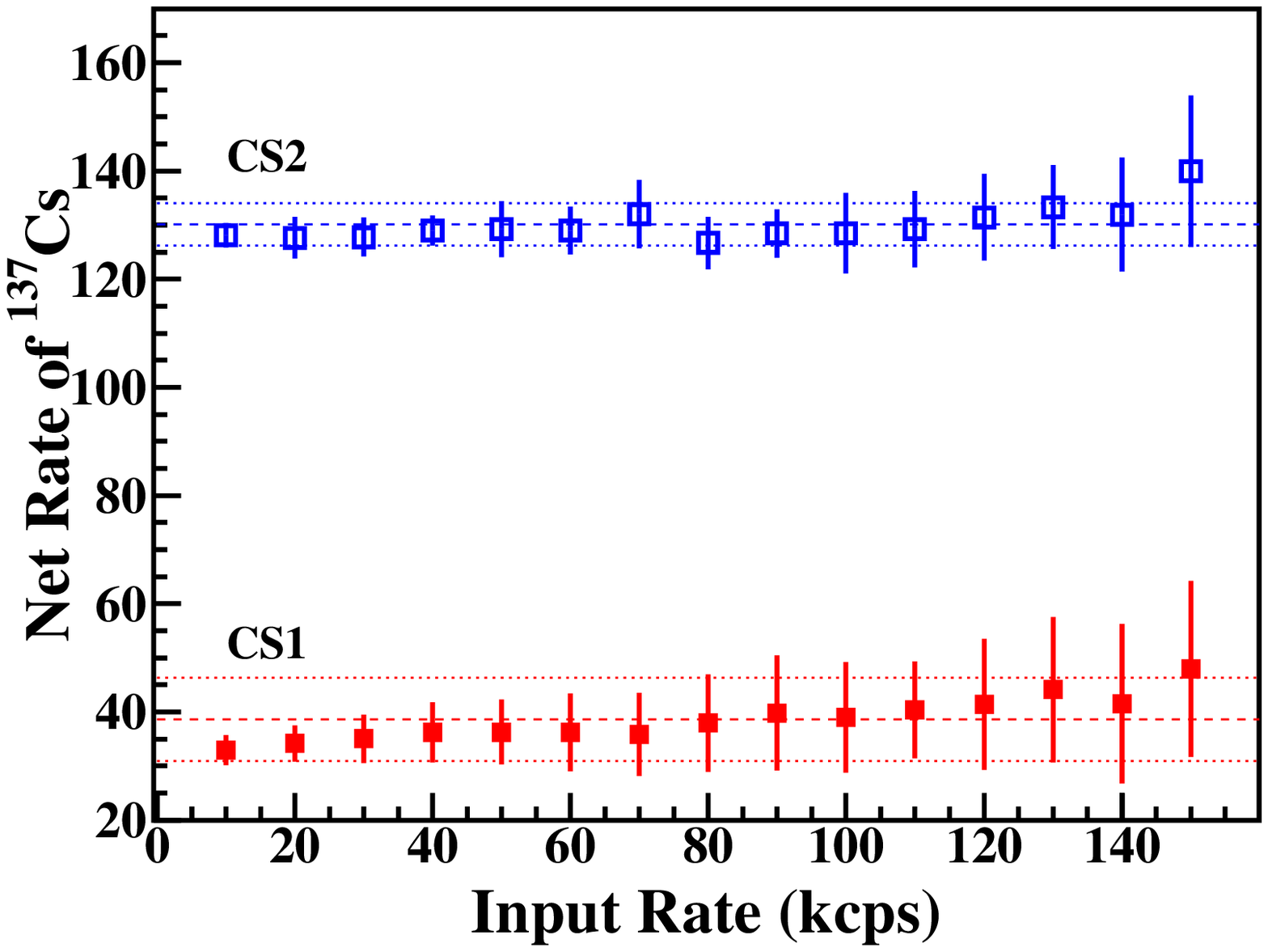}
\caption{(Color online) The net rate of the \cs137 measured at
different \co60 irradiation rate for CS1 and CS2.  The error bars
denote the standard deviation of the net rate distribution in the
repeated measurements.} \label{nrate1}
\end{figure}

Figure \ref {nrate1} further shows the profile of the measured net
counting rate of \cs137 as a function of the input counting rate
 for CS1 and CS2.  It is shown that below 100 \kcps
input rate, the net rate keeps constant, revealing a reliable
extraction of the net counting rate of the EOI \cs137 . Again for
CS2, corresponding to a deep burnup element in our design, the
standard variance of the averaged net rate is within 3\%.

\section {Summary}
In summary, we have studied experimentally the feasibility of
utilizing an HPGe detector in the determination of the fuel element
burnup in a future bed-like reactor. In the multi source
measurement, the passthrough curves are measured by varying the
shaping and flattop time of the spectrometer. The plateau of the
passthrough rate covers the radiation background level from 100 to
200 \kcps with adequate energy resolution and allows to implement
the activity measurement of the burnup indicator \cs137 within half
minute. The energy resolution, reflected by the FWHM of the full
energy peak of $^{137}$Cs in this measurement, allows to identify
the peak of $^{137}$Cs with its various neighboring $\gamma$ rays
despite a slight degradation of 20\% at very high background level.
The precision ($1\sigma$) of the net counting rate  of the EOI
$^{137}$Cs under various conditions is studied. It is demonstrated
that the subtraction of the background contributes significantly to
the total precision of the net area of $^{137}$Cs peak. By fine
tuning the background window and fixing the range of interest, the
precision ($1\sigma$) is optimized to 2.8\% with the typical
incident radiation rate and $^{137}$Cs intensity that is relevant to
the deep burnup situation in the future application. It is worth
mentioning that because the background composition is much more
complicated in the real reactor application than the experimental
conditions presented here, a certain deterioration of the precision
is expected and demands a detailed simulation (via MCNP or Geant4
for instance) of the real background irradiation.

\vspace{-1mm}
\centerline{\rule{80mm}{0.1pt}}

\clearpage

\end{document}